%% file: Het_in_MARL.tex
\PassOptionsToPackage{dvipsnames}{xcolor}

\documentclass[sigconf,colorlinks=true, citecolor=blue!70!black, linkcolor=blue!70!black, urlcolor=blue!70!black, filecolor=blue!70!black]{aamas} 

\makeatletter
\def\@acmBadgeL@image{}
\def\@acmBadgeR@image{}
\def\@acmBadgeL@url{}
\def\@acmBadgeR@url{}
\makeatother

\usepackage{balance} 

\usepackage[utf8]{inputenc} 
\usepackage[T1]{fontenc}    
\usepackage{url}            
\usepackage{booktabs}       
\usepackage{amsmath}        

\usepackage{amssymb}        
\usepackage{nicefrac}       
\usepackage{microtype}      
\usepackage{xcolor}         
\usepackage{graphicx,subfigure}
\usepackage{bm}
\usepackage{multicol}
\usepackage{wrapfig}
\usepackage[most]{tcolorbox}
\usepackage{algorithm}
\usepackage{algorithmicx}
\usepackage{algpseudocode}
\algrenewcommand\algorithmicrequire{\textbf{Input:}}
\algrenewcommand\algorithmicensure{\textbf{Output:}}
\algrenewcommand\algorithmiccomment[1]{\textcolor{blue!60!black}{// #1}}
\algrenewcommand\algorithmicend{\textbf{end}}
\algrenewcommand\algorithmicif{\textcolor{red!70!black}{\textbf{if}}}
\algrenewcommand\algorithmicthen{\textcolor{red!70!black}{\textbf{then}}}
\algrenewcommand\algorithmicelse{\textcolor{red!70!black}{\textbf{else}}}
\algrenewcommand\algorithmicfor{\textcolor{green!60!black}{\textbf{for}}}
\algrenewcommand\algorithmicforall{\textcolor{green!60!black}{\textbf{for all}}}
\algrenewcommand\algorithmicloop{\textcolor{green!60!black}{\textbf{loop}}}
\algrenewcommand\algorithmicwhile{\textcolor{green!60!black}{\textbf{while}}}
\algrenewcommand\algorithmicrepeat{\textcolor{green!60!black}{\textbf{repeat}}}
\algrenewcommand\algorithmicuntil{\textcolor{green!60!black}{\textbf{until}}}
\algrenewcommand\algorithmicprocedure{\textcolor{purple!70!black}{\textbf{procedure}}}
\algrenewcommand\algorithmicfunction{\textcolor{purple!70!black}{\textbf{function}}}
\algrenewcommand\algorithmicreturn{\textcolor{orange!70!black}{\textbf{return}}}
\usepackage{xcolor}
\hypersetup{
    colorlinks=true,
    citecolor=blue!80!black,
    linkcolor=blue!80!black,
    urlcolor=cyan!70!black,
    filecolor=magenta!70!black,
    anchorcolor=red!70!black,
    menucolor=blue!80!black,
    pdfborder={0 0 0},
    pdfstartview=FitH,
    pdfpagemode=UseOutlines,
    bookmarksopen=true,
    bookmarksopenlevel=2,
    linktocpage=true,
    pdfhighlight=/O,
    pdfauthor={Tianyi Hu, Zhiqiang Pu, Yuan Wang, Tenghai Qiu, Min Chen, Xin Yu},
    pdftitle={Heterogeneity in Multi-Agent Reinforcement Learning},
    pdfsubject={Multi-Agent Reinforcement Learning, Heterogeneity},
    pdfkeywords={MARL, Heterogeneity, Policy Diversity}
}
\usepackage{tabularx}

\usepackage{pifont}
\usepackage{multirow}
\usepackage{color} %
\usepackage{booktabs}
\usepackage{rotating}

\setcopyright{ifaamas}
\acmConference[AAMAS '26]{Proc.\@ of the 25th International Conference
on Autonomous Agents and Multiagent Systems (AAMAS 2026)}{May 25 -- 29, 2026}
{Paphos, Cyprus}{C.~Amato, L.~Dennis, V.~Mascardi, J.~Thangarajah (eds.)}
\copyrightyear{2026}
\acmYear{2026}
\acmDOI{}
\acmPrice{}
\acmISBN{}

\acmSubmissionID{<<1592>>}

\title[Heterogeneity in MARL]{Heterogeneity in Multi-Agent Reinforcement Learning}

\author[Tianyi Hu, Zhiqiang Pu, Yuan Wang, Tenghai Qiu, Min Chen, Xin Yu]{Tianyi Hu\textsuperscript{1,2,3}, Zhiqiang Pu\textsuperscript{1,2,3}, Yuan Wang\textsuperscript{1,2,3}, Tenghai Qiu\textsuperscript{1,2}, Min Chen\textsuperscript{1,2}, Xin Yu\textsuperscript{1,2}}
\affiliation{
  \institution{\textsuperscript{\rm 1}Institute of Automation, Chinese Academy of Sciences\\
    \textsuperscript{\rm 2}National Key Laboratory of Cognition and Decision Intelligence for Complex Systems\\
    \textsuperscript{\rm 3}School of Artificial Intelligence, University of Chinese Academy of Sciences}
\country{Beijing, China}
  }

\begin{abstract}

\textit{Heterogeneity} is a fundamental property in multi-agent reinforcement learning (MARL), which is closely related not only to the functional differences of agents, but also to policy diversity and environmental interactions. However, the MARL field currently lacks a rigorous definition and deeper understanding of heterogeneity. This paper systematically discusses heterogeneity in MARL from the perspectives of \textit{definition}, \textit{quantification}, and \textit{utilization}.
First, based on an agent-level modeling of MARL, we categorize heterogeneity into five types and provide mathematical definitions.
Second, we define the concept of heterogeneity distance and propose a practical quantification method.
Third, we design a heterogeneity-based multi-agent dynamic parameter sharing algorithm as an example of the application of our methodology.
Case studies demonstrate that our method can effectively identify and quantify various types of agent heterogeneity. Experimental results show that the proposed algorithm, compared to other parameter sharing baselines, has better interpretability and stronger adaptability.
The proposed methodology will help the MARL community gain a more comprehensive and profound understanding of heterogeneity, and further promote the development of practical algorithms.\footnotemark[1]

\end{abstract}


\newcommand{\BibTeX}{\rm B\kern-.05em{\sc i\kern-.025em b}\kern-.08em\TeX}

\newcommand{\secref}[1]{\hyperref[#1]{Section~\ref*{#1}}}
\newcommand{\figref}[1]{\hyperref[#1]{Figure~\ref*{#1}}}
\newcommand{\tabref}[1]{\hyperref[#1]{Table~\ref*{#1}}}
\providecommand{\algoref}[1]{\hyperref[#1]{Algorithm~\ref*{#1}}}

\begin{document}


\pagestyle{fancy}
\fancyhead{}


\maketitle 

\footnotetext[1]{Please feel free to contact us: hutianyi2021@ia.ac.cn (First Author); zhiqiang.pu@ia.ac.cn (Corresponding Author). Code is available at \url{https://github.com/Harry67Hu/HetDPS}.}


\input{MainTex/01-Intro}

\input{MainTex/02-Pre}
\input{MainTex/03-Het}
\input{MainTex/04-MetaHet}

\input{MainTex/05-HetDPS}

\input{MainTex/06-EXP}

\input{MainTex/07-Conclusion}

\input{MainTex/Appendix}

\newpage
\bibliographystyle{ACM-Reference-Format} 
\bibliography{Het_in_MARL}


\end{document}

%% file: MainTex/01-intro.tex
\section{Introduction}

Multi-agent reinforcement learning (MARL) has achieved success in various real-world applications, such as swarm robotic control~\citep{swarmcontrol}, autonomous driving~\citep{autonomousdriving}, and large language model fine-tuning~\citep{CORY}. However, most MARL studies focus on policy learning for homogeneous multi-agent systems (MAS), overlooking in-depth discussions of heterogeneous multi-agent scenarios~\citep{ning2024survey}. \textit{Heterogeneity} is a common phenomenon in multi-agent systems. For example, in nature, different species of fish collaborate to find food~\citep{shrimp}; in human society, diverse teams demonstrate higher intelligence and resilience~\citep{dall2013distributed,young1993evolution}; and in artificial systems, aerial drones and ground vehicles cooperate to monitor forest fires~\citep{lwowski2017task}. Heterogeneity can enhance system functionality, reduce costs, and improve robustness, but effectively leveraging heterogeneity remains a key challenge in multi-agent system~\citep{heterogeneity}. As an approach of learning through environmental interactions, MARL can effectively enable multi-agent systems to learn collaborative policies. Hence, exploring heterogeneity from a reinforcement learning perspective would significantly broaden the applicability of MARL.

In the current MARL field, although some works explicitly or implicitly mention agent heterogeneity, only a few focus on its definition and identification. Regarding explicit discussion of heterogeneity, studies have explored communication issues~\citep{HetNet}, credit assignment~\citep{GHQ}, and zero-shot generalization~\citep{HetZero} in heterogeneous MARL. However, these works limit their focus to agents with clear functional differences and lack definitions of agent heterogeneity. On the other hand, many studies explore policy diversity in MARL. Some encourage agents to learn distinguishable behaviors based on identity or trajectory information~\citep{EOI, CDS}, some works group agents using specific metrics~\citep{RODE, SePS}, and some quantify policy differences~\citep{SND, MADPS} and design algorithms to control policy diversity~\citep{DiCo}.
\begin{figure}[h]
  \centering
  \includegraphics[width=0.75\linewidth]{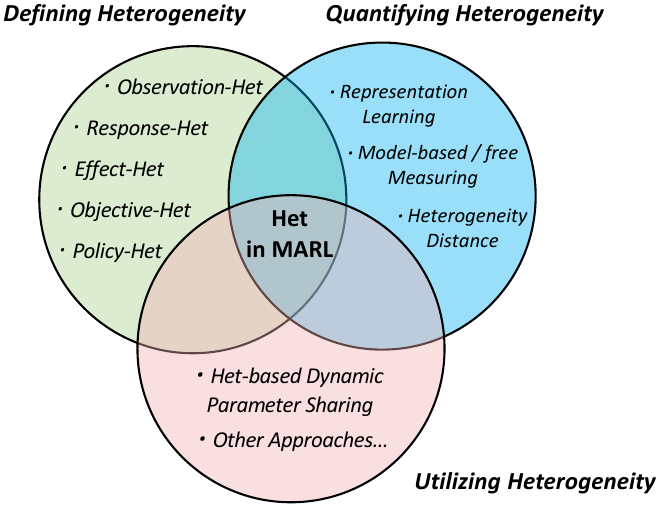}
  \caption{Our Philosophy. We aim to systematically discuss heterogeneity in MARL, establishing methodologies for defining, quantifying and  utilizing heterogeneity.}
  \label{fig:Philosophy}
  \Description{Illustration of the research philosophy showing three main components: defining heterogeneity through agent-level modeling, quantifying heterogeneity using distance metrics, and utilizing heterogeneity in MARL algorithms.}
  \vspace{-5pt}
\end{figure}
However, these works do not adequately address where policy diversity originates or how it fundamentally relates to agent differences. In terms of defining and classifying heterogeneity in MARL, \citep{HetGPPO} divides heterogeneity into physical and behavioral types but lacks a mathematical definition. \citep{HetNet} provides extended POMDP for heterogeneous MARL settings, but do not classify or define heterogeneity. Others introduce the concept of local transition heterogeneity~\citep{GHQ}, but does not cover all elements of MARL. 
Currently, there is still a lack of \textit{systematic analysis of agent heterogeneity from the MARL perspective}.
To fill the aforementioned gaps, we conduct a series of studies on defining, quantifying, and utilizing heterogeneity in the MARL domain, the philosophy of our study can be found in Figure~\ref{fig:Philosophy}. And more details of related work can be found in Appendix A. Our contributions are summarized as follows:

\noindent\textbullet~\textbf{Defining Heterogeneity:} Based on an agent-level model of MARL, we categorize heterogeneity into observation heterogeneity, response transition heterogeneity, effect transition heterogeneity, objective heterogeneity, and policy heterogeneity, and provide corresponding definitions.

\noindent\textbullet~\textbf{Quantifying Heterogeneity:} We define the heterogeneity distance, and propose a quantification method based on representation learning, applicable to both model-free and model-based settings. Additionally, we give the concept of meta-transition heterogeneity to quantify agents’ comprehensive heterogeneity.

\noindent\textbullet~\textbf{Utilizing Heterogeneity:} We develop a multi-agent dynamic parameter-sharing algorithm based on heterogeneity quantification, which offers better interpretability and fewer task-specific hyperparameters compared to other related parameter-sharing methods.

In this paper, we adopt a discussion approach that progresses \textit{from theory to practice} and \textit{from general to specific}. The overall structure is organized as follows: Section~\ref{sec:Preliminaries} introduces the agent-level modeling of the MARL primal problem; Section~\ref{sec:Definition} provides the classification and definition of heterogeneity in MARL; Section~\ref{sec:Quantifying} proposes the method for quantifying heterogeneity and presents case studies; Section~\ref{sec:Utilizing} describes the dynamic parameter-sharing algorithm; Section~\ref{sec:exp} provides the related experimental results; and Section~\ref{sec:conclusion} summarizes the paper.

%% file: MainTex/02-Pre.tex
\section{Preliminaries}
\label{sec:Preliminaries}

\textbf{Primal Problem of MARL.} 
In this paper, we use Partially Observable Markov Game (POMG)~\citep{POMG1, POMG2} as the general model for the primal problem of MARL.\footnote{POMG is an extension of POMDP for multi-agent settings, with the basic extension path being MDP $\to$ POMDP $\to$ POMG~\citep{POMGs}. 
Please refer to Appendix D to see a more detailed explanation of POMG.} To better study agent heterogeneity, we adopt an agent-level modeling approach similar to that in~\citep{HetNet, MADRLSurvey}. A POMG is defined as an 8-tuple, represented as follows:
\begin{equation}
\langle N, \{S^i\}_{i\in N}, \{O^i\}_{i\in N}, \{A^i\}_{i\in N}, \{\Omega^i\}_{i\in N}, \{\mathcal{T}^i\}_{i\in N}, \{r_i\}_{i\in N}, \gamma \rangle,
\label{eq:POMG}
\end{equation}

Among all elements in Expression~\ref{eq:POMG}, $N$ is the set of all agents, $\{S^i\}_{i\in N}$ is the global state space which can be factored as $\{S^i\}_{i\in N} =\times_{i\in N} S^{i} \times S^{E}$, where $S^{i}$ is the state space of an agent $i$, and $S^{E}$ is the environmental state space, corresponding to all the non-agent components. $\{O^i\}_{i\in N}=\times_{i\in N} O^{i}$ is the joint observation space 
and $\{A^i\}_{i\in N}=\times_{i\in N} A^{i}$ is the joint action space of all agents. $\{\Omega^i\}_{i\in N}$ is the set of observation functions.  $\{\mathcal{T}^i\}_{i\in N}=(\mathcal{T}^1, \cdots, \mathcal{T}^{|N|},\mathcal{T}^E)$ is the collection of all agents' transitions and the environmental transition. Finally,  $\{r_i\}_{i\in N}$ is the set of reward functions of all agents and $\gamma$ is the discount factor.

Here, we give the independent and dependent variables for each function and their notation. At each time step $t$, an agent $i$ receives an observation $o^i_t \sim \Omega^{i}(\cdot|\hat{s}_t)$, where $\hat{s}_t \in \{S^i\}_{i\in N}$ is the global state at time $t$. Then, agent $i$ makes a decision based on its observation, resulting in an action $a^i_t \sim \pi_i(\cdot|o^i_t)$. The environment then collects actions from all agents to form the global action $\hat{a}_t = (a^1_t, \dots, a^{|N|}_t)$. We assume that the local state transition of agent $i$ is influenced by the global state and global action, so its local state transitions to a new state $s^i_{t+1} \sim \mathcal{T}^i(\cdot|\hat{s}_t, \hat{a}_t)$. Similarly, the states of other agents and the environment also transition, yielding the next global state $\hat{s}_{t+1} = (s^1_{t+1}, \dots, s^{|N|}_{t+1}, s^E_{t+1}) \sim (\mathcal{T}^1(\cdot|\hat{s}_t, \hat{a}_t), \dots, \mathcal{T}^{|N|}(\cdot|\hat{s}_t, \hat{a}_t), \mathcal{T}^E(\cdot|\hat{s}_t, \hat{a}_t)) = \{\mathcal{T}^i\}_{i\in N}(\cdot|\hat{s}_t, \hat{a}_t)$.
At the same time, all agents receive rewards, with the reward for a specific agent $i$ given by $r^i_t \sim r^{i}(\cdot|\hat{s}_t, \hat{a}_t)$.

The objective of MARL is to solve POMG by finding an optimal joint policy that maximizes the cumulative reward for all agents. We denote the individual optimal policy for agent $i$ as $\pi_i^{*}$ and the optimal joint policy as $\hat{\pi}^{*}$, which can be expressed as $\hat{\pi}^{*}=(\pi_1^{*}, \dots, \pi_{|N|}^{*})$. The optimal joint policy for a POMG can be obtained through the following equation:

\begin{equation}
\pi_i^{*} = \arg\max_{\hat{\pi}} \mathbb{E}_{\hat{\pi}} \left[ \sum_{k=0}^{\infty} \gamma^k \sum_{i\in N} r^i_{t+k} \Big| \hat{s}_t = \hat{s}_0 \right],
\label{eq:joint policy}
\end{equation}
where $\gamma$ is the discount factor, and the expectation is taken over the trajectories via joint policy $\hat{\pi}$ starting from the initial state $\hat{s}_0$.

%% file: MainTex/03-Het.tex
\section{Taxonomy and Definition of Heterogeneity in MARL}
\label{sec:Definition}

\begin{table*}[ht]
\caption{Five Types of Heterogeneity in MARL}
\centering
\small
\begin{tabular}{p{3cm}p{4cm}p{3.5cm}p{5cm}}
\toprule
\textbf{Heterogeneity Type} & \textbf{Heterogeneity Description} & \textbf{Related POMG Elements} & \textbf{Mathematical Definition} \\
\midrule
\textbf{\textit{Observation}} \newline \textbf{\textit{Heterogeneity}} & Describes the differences of agents in observing global information & Agent's observation space and observation function & Agents $i$ and $j$ are observation heterogeneous if: \ding{192} $O^{i} \neq O^{j}$; or \ding{193} $\exists \hat{s} \in \{S^i\}_{i \in N}$, $\Omega^{i}(\cdot|\hat{s}) \neq \Omega^{j}(\cdot|\hat{s})$ \\
\midrule
\textbf{\textit{Response Transition}} \newline \textbf{\textit{Heterogeneity}} & Describes the differences of agents in how their state transitions are affected by global environmental components (\textit{environment-to-self}) & Agent's state space and local state transition function & Agents $i$ and $j$ are response transition heterogeneous if: \ding{192} $S^{i} \neq S^{j}$; or \ding{193} $\exists \hat{s} \in \{S^i\}_{i \in N}$, $\hat{a} \in \{A^i\}_{i \in N}$, $\mathcal{T}^i(\cdot|\hat{s}, \hat{a}) \neq \mathcal{T}^j(\cdot|\hat{s}, \hat{a})$ \\
\midrule
\textbf{\textit{Effect Transition}} \newline \textbf{\textit{Heterogeneity}} & Describes the differences of agents in how their states and actions impact global state transitions (\textit{self-to-environment}) & Agent's action space, state space, and global state transition function & Agents $i$ and $j$ are effect transition heterogeneous if: \ding{192} $S^{i} \neq S^{j}$; or \ding{193} $A^{i} \neq A^{j}$; or \ding{194} $\exists s' \in S^{-i}, a' \in A^{-i}, s \in S^{i}, a \in A^{i}$, $\mathcal{T}^{-i}(\cdot|s', s, a', a) \neq \mathcal{T}^{-j}(\cdot|s', s, a', a)$ \\
\midrule
\textbf{\textit{Objective}} \newline \textbf{\textit{Heterogeneity}} & Describes the differences of agents in the objective they aim to achieve & Agent's reward function & Agents $i$ and $j$ are objective heterogeneous if: \ding{192} $\exists \hat{s} \in \{S^i\}_{i \in N}$, $\hat{a} \in \{A^i\}_{i \in N}$, $r^i(\cdot|\hat{s}, \hat{a}) \neq r^j(\cdot|\hat{s}, \hat{a})$ \\
\midrule
\textbf{\textit{Policy}} \newline \textbf{\textit{Heterogeneity}} & Describes the differences of agents in their decision-making based on observations & Agent's observation space, action space, and policy & Agents $i$ and $j$ are policy heterogeneous if: \ding{192} $O^{i} \neq O^{j}$; or \ding{193} $A^{i} \neq A^{j}$; or \ding{194} $\exists o \in O^{i}$, $\pi_i(\cdot|o) \neq \pi_j(\cdot|o)$ \\
\bottomrule
\end{tabular}
\label{tab:heterogeneity}
\end{table*}

\textbf{Heterogeneity in MAS.}  
Our goal is to define agent heterogeneity from the perspective of MARL. Before achieving this, we discuss heterogeneity in MAS across various disciplines. Early studies~\citep{dudek1996taxonomy,parker2000lifelong} define heterogeneity as differences in \textit{physical structure} or \textit{functionality} of agents, which aligns with common understanding. Later work~\citep{panait2005cooperative} describes heterogeneity as differences in agent \textit{behavior}, further expanding its meaning. 
Recently,~\citep{heterogeneity} points out that heterogeneity may be a complex phenomenon, related not only to the \textit{inherent properties} of agents, but also to their \textit{interactions with environment}. Thus, heterogeneity in MARL should not be limited to inherent functional differences of agents, but should also fully consider various coupling effects of agents within the environment.

\textbf{Heterogeneity in MARL.}
The fundamental modeling of MARL primal problem provides convenience for defining heterogeneity. This modeling specifies all MARL elements, delineating the boundaries of the problem discussion~\footnote{In this paper, we focus on the heterogeneity of MARL under the conventional POMG problem. Additional discussions on unconventional heterogeneity types are provided in Appendix E.} and ensuring the completeness of the discussion.

We focus on the heterogeneity \textit{among agents} within a same POMG. As mentioned in the previous sections, functions in POMG can serve as bridges linking other elements. Therefore, we focus on the functions and classify heterogeneity into five types. This approach can avoid redundant classification, and ensure coverage of each agent-level element. Specifically, these five types of heterogeneity are: \textit{Observation heterogeneity}, \textit{Response transition heterogeneity}, \textit{Effect transition heterogeneity}, \textit{Objective heterogeneity}, and \textit{Policy heterogeneity}. Their specific descriptions and definitions are given in Table~\ref{tab:heterogeneity}. In this table, $S^{-i} = \times_{k \in N, k \neq i} S^{k} \times S^{E}$ represents the joint state space of all agents except agent $i$, reflecting the influence of the agent on other states. Similarly, $A^{-i}$ denotes the joint action space excluding agent $i$, and $\mathcal{T}^{-i}$ is the collection of state transitions excluding agent $i$.

The definitions in this section are relatively straightforward: if there are any differences in the associated elements, the agents are considered heterogeneous. We need to emphasize that our work goes beyond this. The quantification methods provided in the next section will be able to characterize the degree of agent heterogeneity related to certain attributes in practical scenarios, which far exceeds the level of definition.

%% file: MainTex/04-MetaHet.tex
\section{Quantifying Heterogeneity in MARL}
\label{sec:Quantifying}

\subsection{Heterogeneity Distance} According to the definition, each type of heterogeneity corresponds to a core function which connects relevant elements in the heterogeneity type. Therefore, we quantify the differences in these core functions to characterize the degree of heterogeneity.\footnote{Quantifying space elements is feasible and even easier to implement. But a space element may appear across multiple heterogeneity types, making it unsuitable as unique identifiers for specific heterogeneity types.} To make the quantification results simpler and more practical, we draw upon the ideas of policy distance from the works \citep{SND} and \citep{MADPS}, and present the concept of heterogeneity distance.

Let the core function corresponding to a certain heterogeneity type $F$ be denoted as $ y \sim F(\cdot|x) $. The formula for calculating the  $F$-heterogeneous distance between two agents $ i $ and $ j $ is given by: 
\begin{equation}
    d_{ij}^F = \int_{x \in X} D[F_i(\cdot|x) \parallel F_j(\cdot|x)] \cdot p(x) \, dx,
\label{eq:distance}
\end{equation}
where $ X $ is the space of independent variables, $ p(x) $ is the probability density function, and $ D [\cdot \parallel \cdot] $ is a measure that quantifies the difference between distributions. 
Unlike the works in \citep{SND} and \citep{MADPS}, we add probability density terms to ensure accuracy and consider the case of multivariate variables.
When the independent variables $x$ consist of multiple factors, the above integral becomes a multivariate integral. Based on Equation~\ref{eq:distance}, we provide the specific expressions for quantifying all heterogeneous distances in Appendix G and discuss the properties of heterogeneous distance in Appendix F.


\begin{figure*}[ht]
  \centering
  \includegraphics[width=0.9\linewidth]{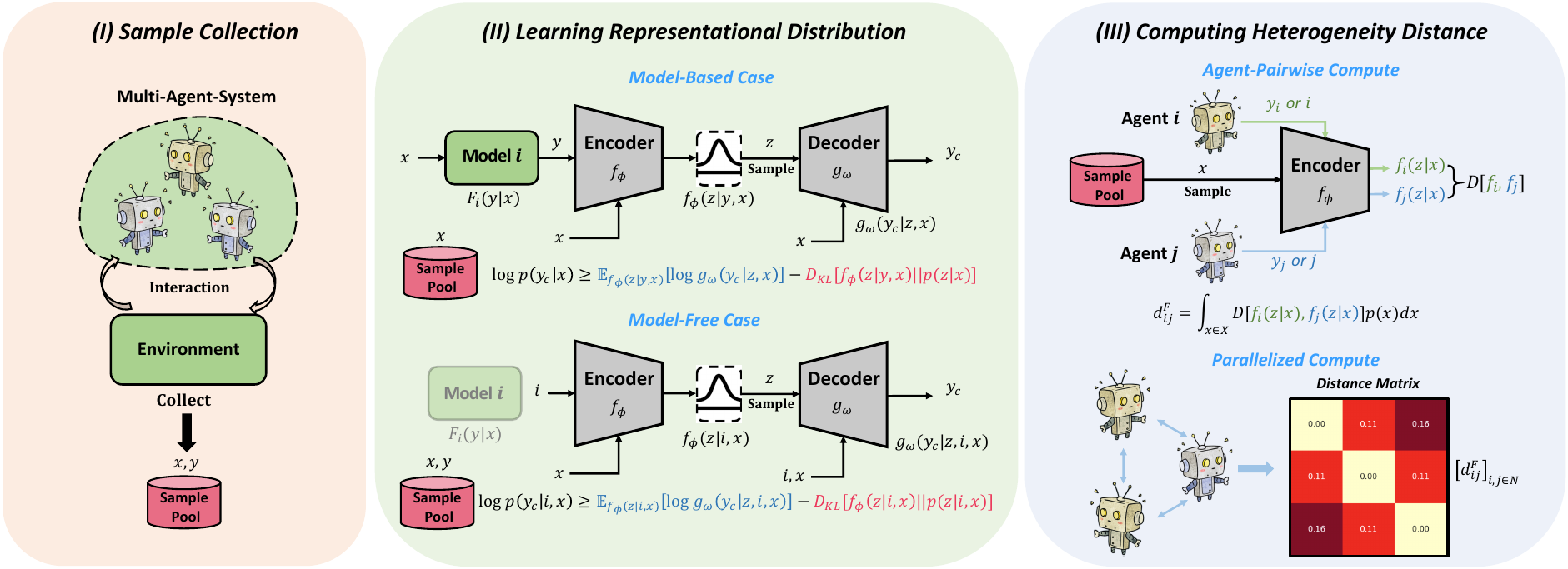}
  \caption{The method of measuring heterogeneity distance based on representation learning.
  }
  \label{fig:het_measure}
  \Description{Flowchart illustrating the three-step process for measuring heterogeneity distance: function extraction, representation learning using conditional variational autoencoder, and distance computation using Wasserstein distance.}
\end{figure*}

\subsection{Practical Method} 

To compute Equation~\ref{eq:distance} in practice, we need to address several core issues: 1. Full space traversal. In practice, it is impossible to traverse the entire space $X$. 2. Measure $D$ is difficult to compute. Even assuming we can obtain model $F$ for each agent, the distribution types of $F$ may vary, making it difficult to calculate measures between different distributions. 3. Handling cases when model $F$ is not available.  More commonly, it is hard to obtain environment-based agent models, especially in practical MARL tasks.

For issue 1, our approach is sampling based on the interaction between agents and the environment. Instead of simply traversing
the space or using random policy exploration, we construct a sample pool using trajectories from the training phase of
MARL. This significantly reduces computational load and filters out
excessive marginal spaces, benefiting the
use of heterogeneity distance in subsequent MARL tasks (Section~\ref{sec:Utilizing}). For issue 2, this has been solved in paper~\citep{MADPS} (computing policy distance). We follow their approach, which performs representation learning on $F$ and maps it to a standardized distribution.
For issue 3, we extend the representation learning method to model-free cases. This helps apply the method in real-world settings and enables us to propose the concept of \textit{Meta-Heterogeneity Distance}. By freely combining different attributes to construct \textit{Meta-Transitions}, the proposed method can quantify the ``comprehensive heterogeneity'' of agents.

Combining these ideas, we propose a practical method as shown in Figure~\ref{fig:het_measure}. 
\textbf{In the first step}, the agents interact with the environment during MARL training to build a sample pool. Notably, the sample pool data is shuffled to ensure that the learned function follows the \textit{Markov} property (independent of historical information).

\textbf{In the second step}, the representational distributions are learned. We discuss this in both model-based and model-free settings, corresponding to cases function $F$ is known and unknown. We adopt the conditional variational autoencoder (CVAE)~\citep{CVAE} for representation learning. In the model-based case, CVAE performs a reconstruction task~\citep{lopez2017conditional}. 
The optimization goal is to maximize the likelihood of the reconstructed variable $\log p(y|x)$. Through derivation, we obtain the evidence lower bound (ELBO) as:
\begin{equation}
\begin{aligned}
    ELBO_{\text{model-based}} = & \mathbb{E}_{f_\phi(z | y, x)}\left[\log g_\omega(y | z, x)\right] \\
    & - D_{KL}\left[f_\phi(z | y, x) \parallel p(z | x)\right],
\end{aligned}
\end{equation}
where $f_\phi$ and $g_\omega$ represent the encoder and decoder, respectively, and $p(z | x)$ is the prior conditional latent distribution. The relevant losses are designed based on ELBO, including a reconstruction loss and a prior-matching loss. 

In the model-free case, CVAE essentially performs a prediction task~\citep{zhang2021conditional}, capturing the model characteristics of each agent. 
The network takes the independent variable $x$ and agent ID $i$ as inputs, using both as conditions to predict $y$. 
The optimization goal is to maximize the likelihood of the predicted $y$ given conditions. Similarly, the corresponding ELBO can be derived as (the derivation for this part can be found in Appendix I):

\begin{equation}
\begin{aligned}
    ELBO_{\text{model-free}} = & \mathbb{E}_{f_\phi(z | i, x)}\left[\log g_\omega(y | z, i, x)\right] \\
    & - D_{KL}\left[f_\phi(z | i, x) \parallel p(z | i, x)\right].
\end{aligned}
\end{equation}

\textbf{In the third step}, the heterogeneity distances for MAS are computed. For each $x$, we obtain the distribution representation using the encoder in either the model-based or model-free manner. The distance under a specific $x$ is computed using the \textit{Wasserstein distance}~\citep{WD} of the prior distribution (\textit{standard Gaussian}). The heterogeneity distance is then calculated via multi-rollout Monte Carlo sampling. In practice, we parallelize this operation~\footnote{Our code is provided in the supplementary material.}, enabling simultaneous computation of distances between all agents on GPUs, significantly improving computational efficiency.

\textbf{Meta-Transition.} The aforementioned method can quantify the heterogeneity of agents for specific types. In practical applications, researchers may also want to quantify the \textbf{comprehensive} heterogeneity of agents to enable operations such as grouping. To this end, we give the \textit{Meta-Transition} model (see Appendix H for details). By measuring the differences between meta-transitions, the comprehensive heterogeneity related to environment can be quantified. We refer to this as the meta-transition heterogeneity distance (Hereafter referred to as \textit{Meta-Het}).


\begin{figure*}[ht]
  \centering
  \includegraphics[width=1.0\linewidth]{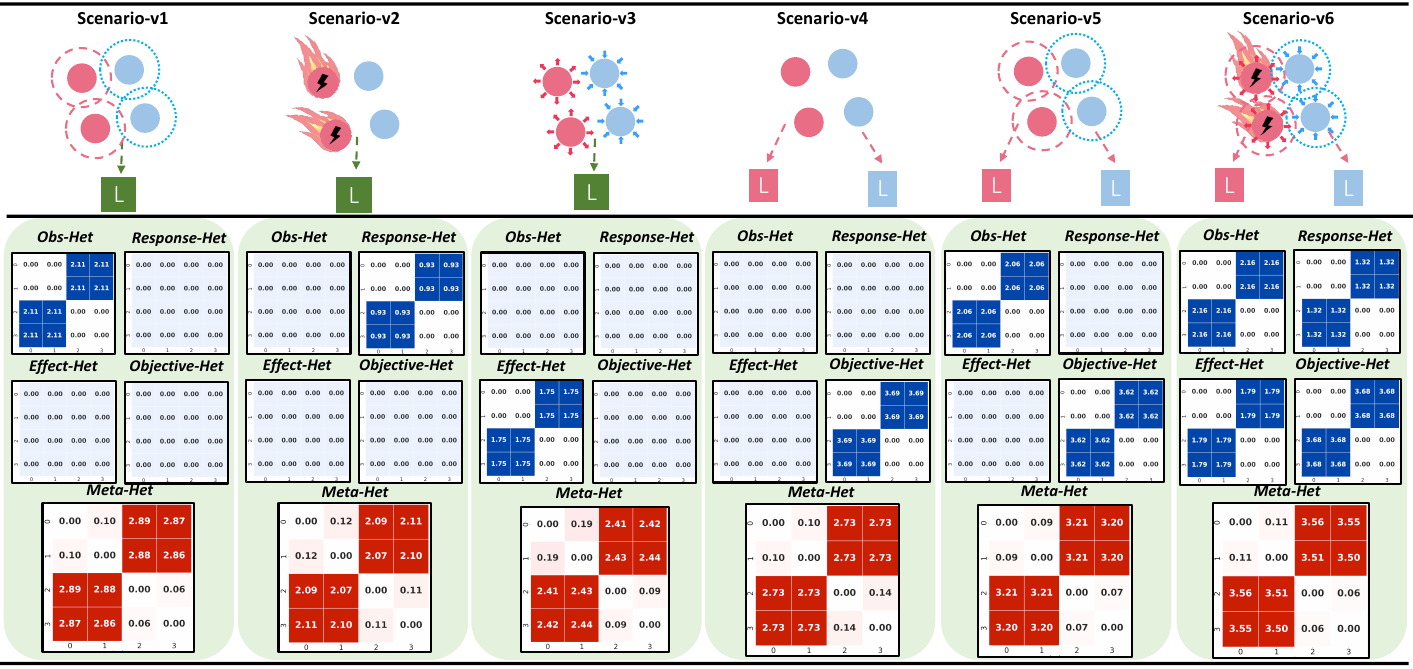}
  \caption{The scenario illustration and heterogeneity distance matrices. In \textit{v1}, the observations of agents from different groups are shuffled in different orders. In \textit{v2}, the max speeds of agents are different. In \textit{v3}, one group of agents applies repulsive force to surrounding entities, while the other attractive force. In \textit{v4}, agents need to move to different landmarks. In \textit{v5}, both the observations and objectives of agents are heterogeneous. In \textit{v6}, all the above properties are heterogeneous. Below each scenario illustration, the corresponding heterogeneity distance matrices are shown. Specifically, \textit{Obs-Het}, \textit{Response-Het}, \textit{Effect-Het}, and \textit{Objective-Het} correspond to observation / response transition / effect transition / objective heterogeneity, respectively.}
  
  \label{fig:case_study_1}
  \Description{Six different multi-agent scenarios (v1-v6) with increasing levels of heterogeneity, each showing scenario illustrations and corresponding heterogeneity distance matrices for observation, response transition, effect transition, and objective heterogeneity types.}
\end{figure*}

\subsection{Case Study}
\label{sec:case_study}

We design a multi-agent spread scenario for case study. In the basic scenario, there are two groups, each with two agents, and their goal is to move to randomly generated landmarks. We create 6 versions of the scenario to show the quantitative results of different types of heterogeneity and \textit{Meta-Het}. As shown in Figure~\ref{fig:case_study_1}, the first 4 versions correspond to the 4 environment-related types of heterogeneity, while the last 2 versions represent cases where multiple types of heterogeneity exist. We use the model-based manner to compute the first four distance matrices, and the model-free manner to compute the \textit{Meta-Het} distance matrix.

The results show that for each type of heterogeneity, our method can accurately capture and identify the differences. And the \textit{Meta-Het} distance between agents in the same group is much smaller than that in different groups. Moreover, as the number of heterogeneity types increases, the \textit{Meta-Het} distance between different groups also increases. These results demonstrate the effectiveness of our method for various environment-related heterogeneities.

\begin{figure}[h]
  \centering
  \includegraphics[width=0.95\linewidth]{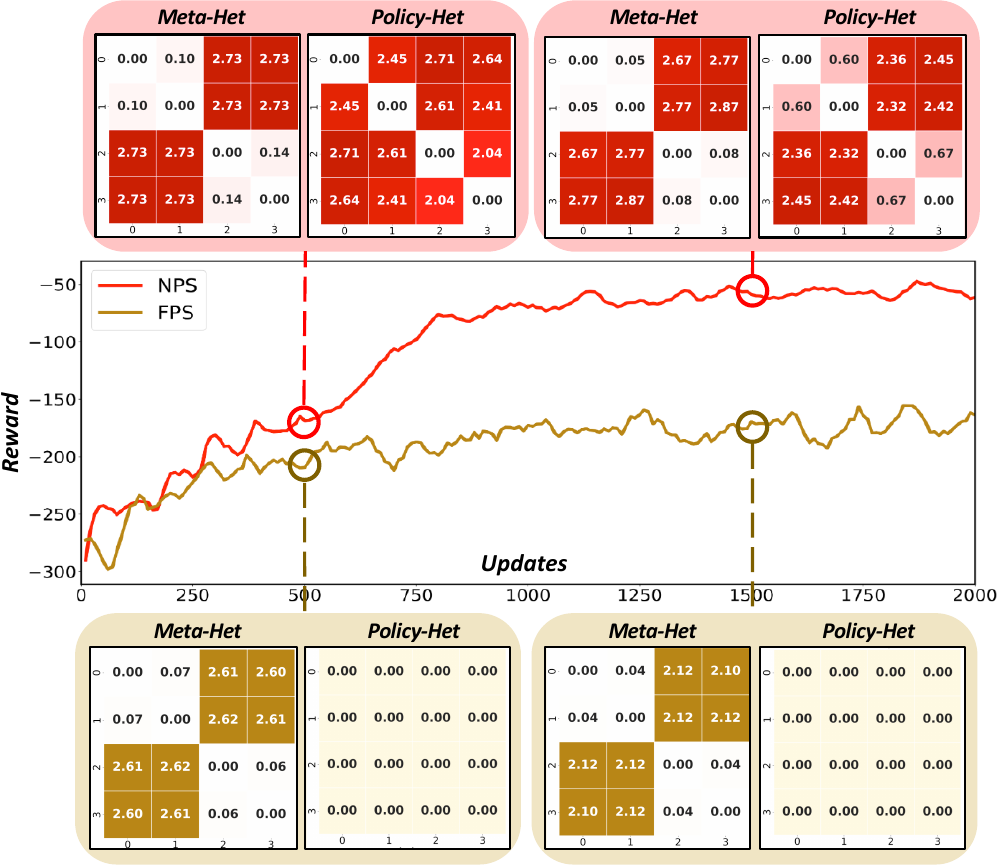}
  \caption{Meta-transition heterogeneity and policy heterogeneity distance matrices during training in our case study.}
  \label{fig:Training_Het}
  \Description{Meta-transition heterogeneity and policy heterogeneity distance matrices showing the evolution of agent differences during training at different update steps for fully parameter sharing and no parameter sharing algorithms.}
\end{figure}

We further quantify the policy heterogeneity distance (\textit{Policy-Het}) and \textit{Meta-Het} distance of agents during the training process. We select two algorithms at the extreme cases of parameter sharing: fully parameter sharing (FPS) and no parameter sharing (NPS) for training in the above scenarios. Figure~\ref{fig:Training_Het} shows the measurement results at 500 and 1500 updates. From the \textit{Policy-Het} results, the policy distance can effectively reveal the evolution of agent policy differences in MARL. From the \textit{Meta-Het} results, the comprehensive agent heterogeneity measurement remains consistent across different learning algorithms, and can identify environmental heterogeneous characteristics in scenarios more rapidly compared to policy evolution.

%% file: MainTex/05-HetDPS.tex
\section{Utilizing Heterogeneity in MARL}
\label{sec:Utilizing}

\begin{table*}[ht]
\centering
\small
\caption{Comparison of different methods and their properties.}
\begin{tabular}{l c c l}
\toprule
\textbf{Method} & \textbf{Paradigm} & \textbf{Adaptive} & \textbf{Relation to Heterogeneity Utilization} \\
\midrule
NPS & No Sharing & No & None \\
FPS & Full Sharing & No & None \\
FPS+id & Full Sharing & No & None \\
Kaleidoscope~\citep{Kaleidoscope} & Partial Sharing & Yes & No utilization, increases agent policy heterogeneity as the bias \\
SePS~\citep{SePS} & Group Sharing & No & Implicitly utilizes objective heterogeneity and response transition heterogeneity \\
AdaPS~\citep{AdaPS} & Group Sharing & Yes & Implicitly utilizes objective heterogeneity and response transition heterogeneity \\
MADPS~\citep{MADPS} & Group Sharing & Yes & Explicitly utilizes policy heterogeneity only \\
HetDPS (ours) & Group Sharing & Yes & Explicitly utilizes heterogeneity, leveraging heterogeneous distance \\
\bottomrule
\end{tabular}
\label{tab:methods_comparison}
\end{table*}

\begin{figure*}[ht]
  \centering
  \includegraphics[width=0.85\linewidth]{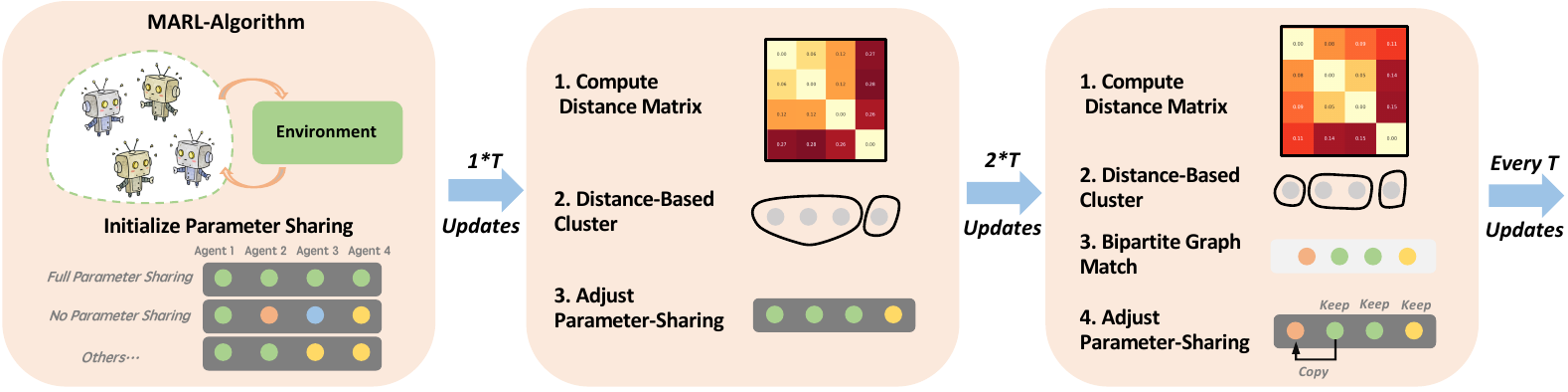}
  \caption{The method of multi-agent dynamic parameter sharing algorithm based on heterogeneity quantification.
  }
  \label{fig:HetDPS}
  \Description{Flowchart of the HetDPS algorithm showing the process of periodic heterogeneity quantification, agent clustering based on heterogeneity distances, and dynamic parameter sharing group assignment.}
\end{figure*}

Based on the case study in Section~\ref{sec:case_study}, the proposed method can not only accurately quantify all types of heterogeneity, but also the ``comprehensive heterogeneity'' among agents. Additionally, the method is independent of the parameter-sharing type used in MARL and can be deployed online, thereby further enhancing its practicality. In this section, we provide a practical application of our methodology to demonstrate its potential in empowering MARL.

We select parameter sharing in MARL as our application context. As a common technique in MARL, parameter sharing can improve sample utilization efficiency~\citep{SNP}, but its excessive use may inhibit agents' policy heterogeneity expression~\citep{MADPS}. Many works have attempted to find a balance between parameter sharing and policy heterogeneity~\citep{Kaleidoscope}. However, existing approaches suffer from two main problems: \textit{poor interpretability}, unable to explain why policy heterogeneity is necessary and to what extent; and \textit{poor adaptability}, manifested by numerous task-specific hyperparameters 

To address these issues, we propose a \textbf{Het}erogeneity-based multi-agent \textbf{D}ynamic \textbf{P}arameter \textbf{S}haring algorithm (HetDPS) with two core ideas (More details can be found in Appendix J):

\textcolor{black}{$\spadesuit$} \textbf{Grouping agents for parameter sharing through heterogeneity distances}. We utilize distance-based clustering methods to group agents, thus avoiding the introduction of task-specific hyperparameters like group number~\citep{SePS,AdaPS} or fusion thresholds~\citep{MADPS}. The heterogeneity distance matrices also enhance the algorithm's interpretability.

\textcolor{black}{$\clubsuit$} \textbf{Periodically quantifying heterogeneity and modifying agents' parameter sharing paradigm}. This approach can help policies escape local optima~\citep{lyle2024normalization}, the effectiveness of such a mechanism has been verified in the MARL domain~\citep{Kaleidoscope}, and even in broader RL areas such as large model fine-tuning~\citep{elastic,CORY}. 

Combining the above ideas, we present the method of HetDPS as illustrated in Figure~\ref{fig:HetDPS}. This approach can be combined with common MARL algorithms and supports various parameter-sharing initialization (e.g., FPS and NPS). After every $T$ updates, the algorithm computes the distance matrix of agents and groups them via distance-based clustering. If clustering exists from the previous cycle, bipartite graph matching is performed between the two clustering results to help agents determine policy inheritance relationships. This \textit{dual-clustering mechanism} effectively enhances the algorithm's adaptability.

We emphasize that utilization of MARL heterogeneity extend beyond this scope. Through our method, researchers can quantify specific types of heterogeneity or composite heterogeneity, which can be integrated with cutting-edge MARL research directions, as detailed in Appendix C.

%% file: MainTex/06-EXP.tex
\begin{table}{}
\centering
\small
\caption{Task information for PMS.}
\begin{tabular}{cc}
\toprule
Task & Agent Type Distribution \\
\midrule
\textit{15a\_3c} & $5-5-5$ \\
\textit{30a\_3c} & $10-10-10$ \\
\textit{15a\_5c} & $3-3-3-3-3$ \\
\textit{30a\_5c} & $3-3-3-12-9$ \\
\bottomrule
\end{tabular}
\label{tab:tasks}
\end{table}

\begin{table}{}
\centering
\small
\caption{Agent distribution in four heterogeneous SMAC tasks.}
\begin{tabular}{cc}
\toprule
Task & Agent Type Distribution \\
\midrule
\textit{3s5z} & 3 Stalkers (0--2) -- 5 Zealots (3--7) \\
\textit{3s5z\_vs\_3s6z} & 3 Stalkers (0--2) -- 5 Zealots (3--7) \\
\textit{MMM} & 2 Marauders (0--1) -- 7 Marines (2--8) -- 1 Medivac (9) \\
\textit{MMM2} & 2 Marauders (0--1) -- 7 Marines (2--8) -- 1 Medivac (9) \\
\bottomrule
\end{tabular}
\label{tab:smac_tasks}
\end{table}

\section{Experiments}
\label{sec:exp}

In this section, we conduct comprehensive comparisons between HetDPS and other parameter sharing methods. Beyond performance comparisons, we also analyze the heterogeneity characteristics of each MARL task with our methodology, to demonstrate the algorithm's interpretability. Additionally, we conduct hyperparameter experiments and efficiency and resource consumption experiments, to show the adaptability and practicality of HetDPS.

\subsection{Experimental Setups}
\label{sec:experimental_setups}

\begin{figure*}[ht]
  \centering
  \includegraphics[width=0.8\linewidth]{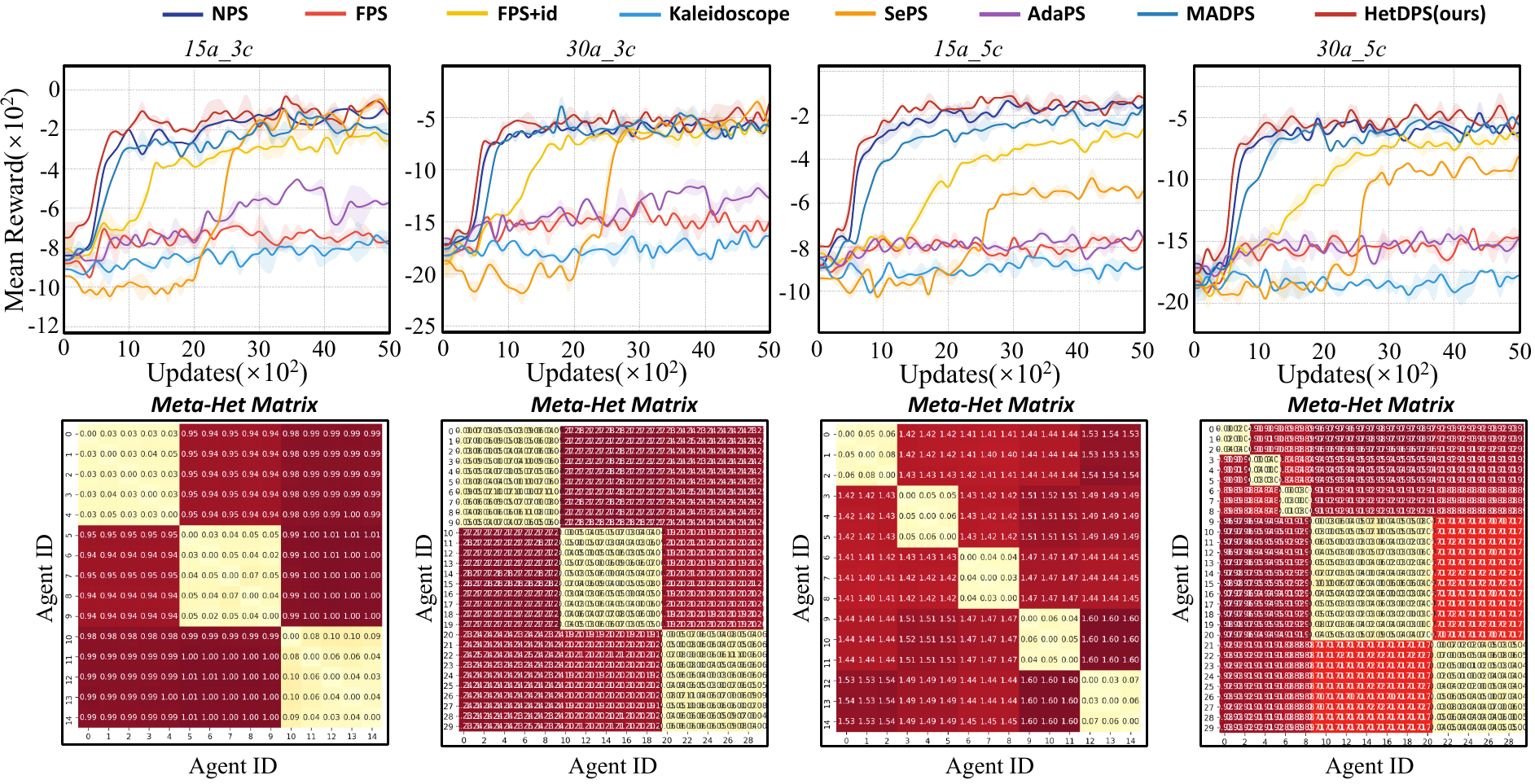}
  \caption{Results on Partical-based Multi-agent Spreading.}
  \label{fig:MPEResult}
  \Description{Training curves and heterogeneity distance matrices for Particle-based Multi-agent Spreading tasks showing performance comparison across different methods.}
\end{figure*}

\textbf{Environments.} \textbf{Partical-based Multi-agent Spreading} \textbf{(PMS)}~\citep{MADPS} is a typical environment in the policy diversity domain. In this environment, multiple agents are randomly generated in the center of the map, while multiple landmarks are generated near the periphery. Both agents and landmarks have various colors, and agents need to move to landmarks with matching colors. Additionally, agents need to form tight formations when they reach the vicinity of landmarks. We employ 4 typical tasks, corresponding to different numbers and color distributions, as detailed in Table~\ref{tab:tasks}. \textbf{The StarCraft Multi-Agent Challenge (SMAC)}~\citep{SMAC} is a popular MARL benchmark, where multiple ally units controlled by MARL algorithms aim to defeat enemy units controlled by built-in AI.

\begin{figure*}[ht]
  \centering
  \includegraphics[width=0.8\linewidth]{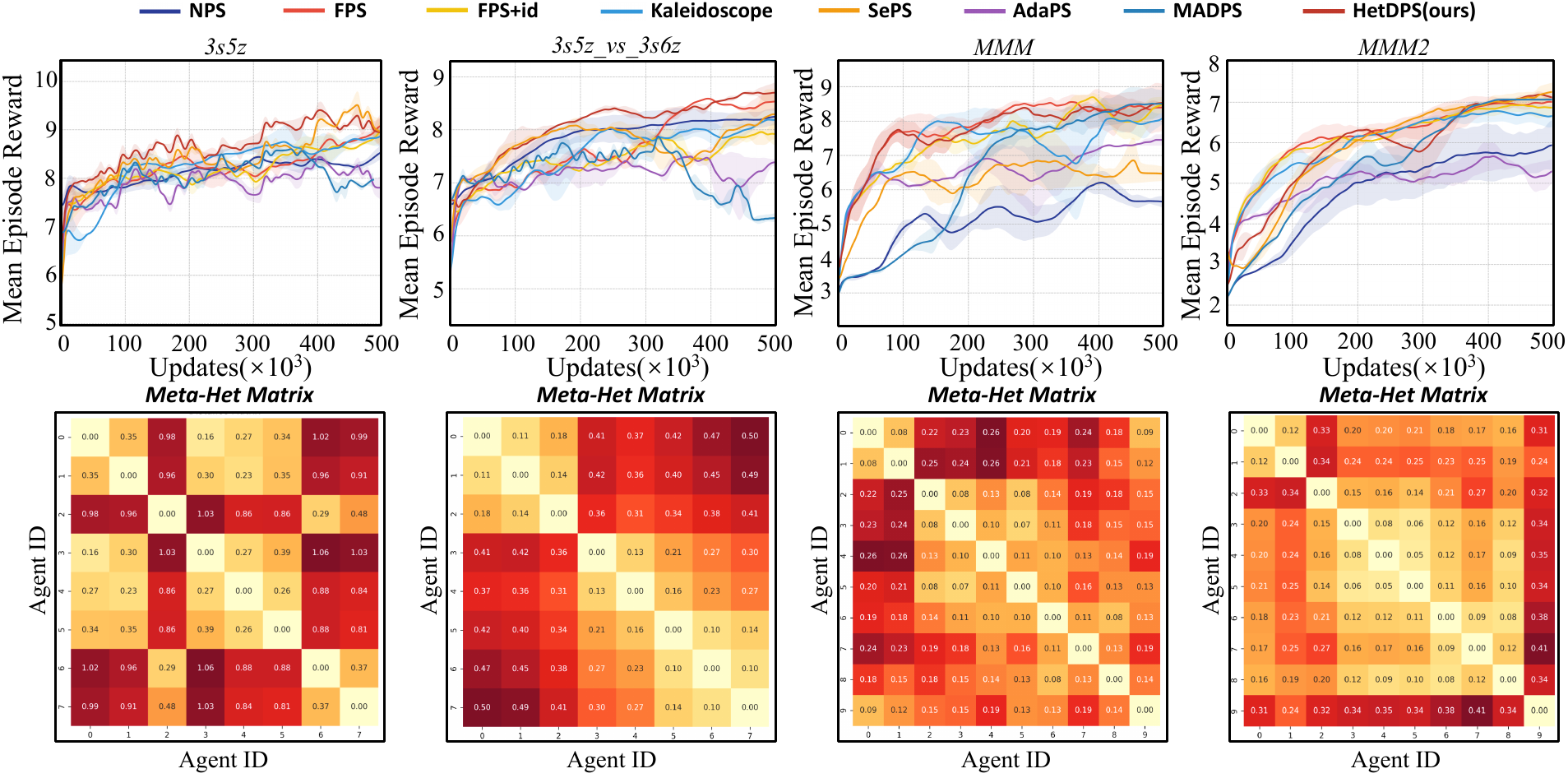}
  \caption{Results on four ``heterogeneous'' tasks in SMAC.}
  \label{fig:SMACHet}
  \Description{Training curves and heterogeneity distance matrices for heterogeneous StarCraft Multi-Agent Challenge tasks showing performance comparison across different methods.}
\end{figure*}

\begin{figure*}[ht]
  \centering
  \includegraphics[width=0.8\linewidth]{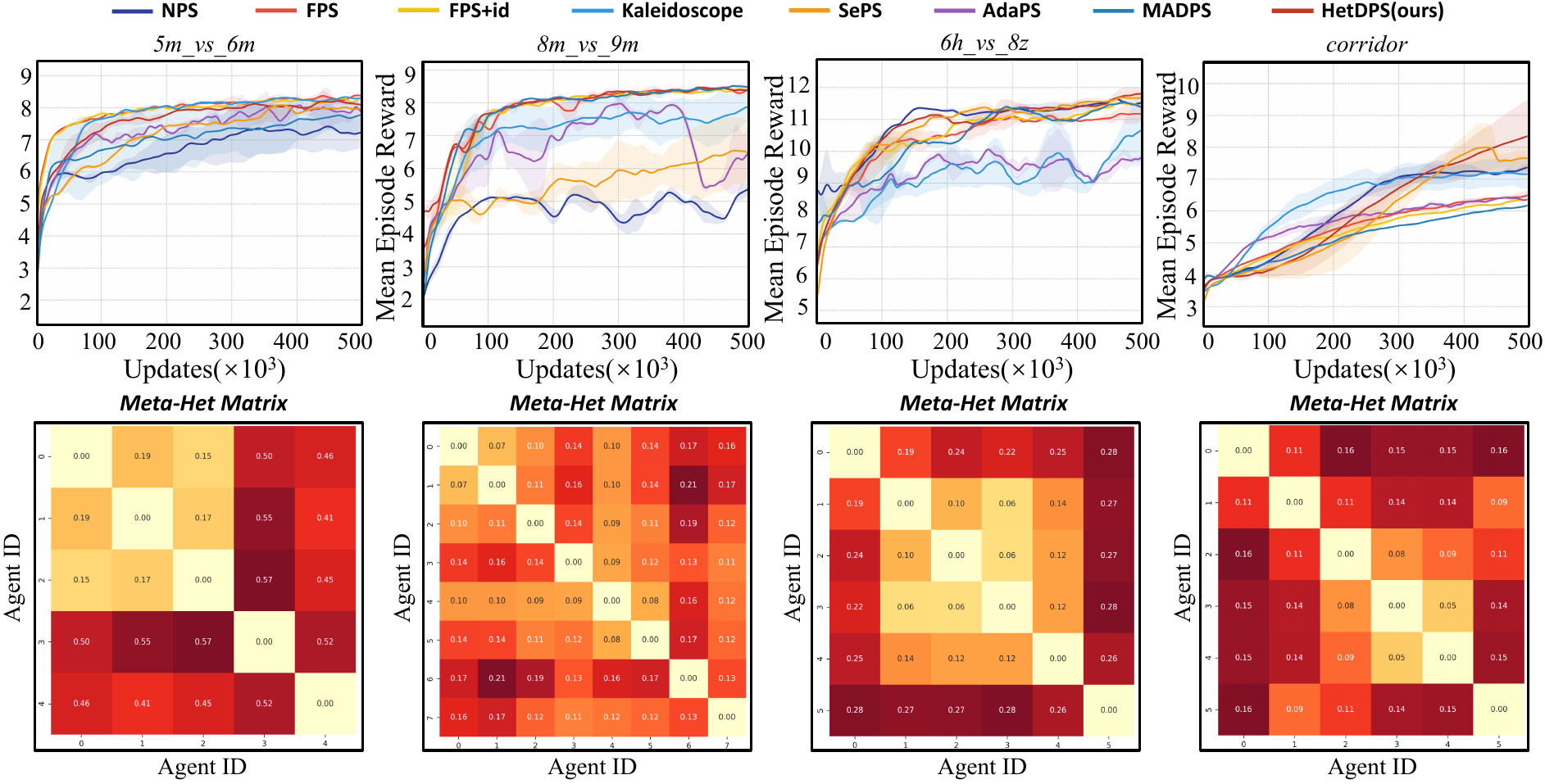}
  \caption{Results on four ``homogeneous'' tasks in SMAC.}
  \label{fig:SMACHom}
  \Description{Training curves and heterogeneity distance matrices for homogeneous StarCraft Multi-Agent Challenge tasks showing performance comparison across different methods.}
\end{figure*}

\textbf{Baselines and training.}
We compare HetDPS with other parameter sharing baselines, as listed in Table~\ref{tab:methods_comparison}. 
As seen from the table, current methods can not effectively utilize heterogeneity. Although some methods implicitly use certain heterogeneity quantification results, the elements they involve are not comprehensive. MADPS, as the only method that explicitly uses policy distance for dynamic grouping, relies on the assumption that policy learning can effectively capture heterogeneity, which lacks practicality. All parameter-sharing methods are integrated with MAPPO, and we use official implementations of the baselines wherever available. For more details of the experiments, see the Appendix K.

\subsection{Results}
\label{sec:Results}

\begin{table*}[htbp]
  \centering
  \small
  \caption{Training efficiency metrics across different methods. Results are normalized with respect to the FPS method.}
  \label{tab:training_efficiency}
  \begin{tabular}{lcccccccc}
    \toprule
    & \textbf{NPS} & \textbf{FPS} & \textbf{FPS+id} & \textbf{Kaleidoscope} & \textbf{SePS} & \textbf{AdaPS} & \textbf{MADPS} & \textbf{HetDPS (ours)} \\
    \midrule
    \textbf{Training Speed} & 0.952x & 1.000x & 0.992x & 0.974x & 0.986x & 0.614x & 0.539x & 0.712x \\
    \bottomrule
  \end{tabular}
  \label{tab:eff}
\end{table*}

\textbf{Performance and interpretability.} The reward curves and corresponding heterogeneity distance matrices are shown in Figure~\ref{fig:MPEResult}, Figure~\ref{fig:SMACHet} and Figure~\ref{fig:SMACHom}. From the reward curve results, we can see that HetDPS achieves either optimal or comparable results across all tasks.

The \textit{Meta-Het} distances in Multi-agent Spreading scenario closely match the type distributions in Table~\ref{tab:tasks}, validating demonstrating the effectiveness of our method in identifying agent heterogeneity. In SMAC, we observe that in simpler tasks like \textit{3s5z} and \textit{MMM}, the heterogeneity distances often do not closely match the original agent types. In \textit{MMM}, agents even tend toward homogeneous policies to improve training efficiency. However, in more difficult tasks such as \textit{3s5z\_vs\_3s6z} and \textit{MMM2}, agents' quantification results closely match their original types for better coordination. This confirms that agent heterogeneity depends on both functional attributes and environment interactions.

Similarly, Figure~\ref{fig:SMACHom} reveals that even ``homogeneous"
agents exhibit emergent heterogeneity from environment interactions, leading to role division. The performance difference between HetDPS and FPS also reflects the impact of role division versus non-division. Our method thus provides both superior performance and strong interpretability for exploring heterogeneity in MARL tasks.

\begin{table}[h]
\centering
\caption{Results of varying quantization intervals in PMS, showing the average rewards of agents.}
\begin{tabular}{lcccc}
\toprule
Quantization Interval & \textit{15a\_3c} & \textit{30a\_3c} & \textit{15a\_5c} & \textit{30a\_5c} \\
\midrule
20 \textit{Updates}  & -10.12 & -300.56 & -50.89 & -350.23 \\
100 \textit{Updates}  & -9.45  & -298.91 & -49.32 & -349.67 \\
200  \textit{Updates} & -10.78 & -301.34 & -51.15 & -351.45 \\
1000 \textit{Updates} & -11.23 & -299.67 & -50.44 & -350.89 \\
2000 \textit{Updates} & -9.87  & -300.12 & -49.78 & -349.12 \\
\bottomrule
\end{tabular}
\label{tab:hyper}
\end{table}

\textbf{Adaptability.} Our approach achieves comparable performance across all tested tasks.
Moreover, we emphasize that \textbf{for all tested tasks, our method uses identical hyperparameters}, without requiring task-specific tuning.
Other baselines require task-specific hyperparameters: e.g., reset interval, reset rate, and diversity loss coefficient for Kaleidoscope; number of clusters and update interval for SePS and AdaPS; fusion/division threshold and quantization interval for MADPS.

HetDPS employs distance-based clustering, eliminating hyperparameters such as cluster number or fusion threshold. Furthermore, by fully accounting for dual-clustering mechanism, HetDPS is insensitive to the quantization interval. Table~\ref{tab:hyper} shows that performance remains stable across quantization intervals ranging from 20 to 2000 in all multi-agent spreading tasks.

\textbf{Cost Analysis.} We conduct an experiment to investigate training efficiency. The experimental results are shown in Table~\ref{tab:eff}. The results indicate that although our method introduces periodic heterogeneity quantification, it does not significantly reduce algorithm efficiency.

\vspace{-5pt}

%% file: MainTex/07-Conclusion.tex
\section{Conclusion}
\label{sec:conclusion}

Heterogeneity manifests in various aspects of MARL. It is not only related to the inherent properties of agents but also to the coupling factors arising from agent-environment interactions. Consequently, agents that appear homogeneous may develop heterogeneity under environmental influences. In this paper, we categorize heterogeneity in MARL into five types and provide definitions. Meanwhile, we propose methods for quantifying these heterogeneities and conduct case studies. Under our theoretical framework, policy diversity is merely a manifestation of policy heterogeneity, fundamentally originating from the division of labor necessitated by agents' environmental heterogeneity (\textit{cause}), serving as an inductive bias (\textit{result}) for solving optimal joint policies. Thus, we introduce the quantification of heterogeneity as prior knowledge into multi-agent parameter-sharing learning, resulting in HetDPS, an algorithm with strong interpretability and adaptability. HetDPS is not the endpoint of our research, but rather a starting point for heterogeneity applications. We believe that by systematically studying the definition, quantification, and application of heterogeneity, future MARL research will more profoundly understand the complex collaboration mechanisms between agents, and pave the way for more intelligent and adaptive collective decision-making systems.

%% file: MainTex/Appendix.tex
\appendix

\section*{Appendix}
\addcontentsline{toc}{section}{Appendix}

\section{Limitations}
\label{app:Limitations}

Although our proposed heterogeneity distance can effectively quantify agent heterogeneity and identify various potential heterogeneities, there remain some limitations in its practical implementation. One limitation is in scaling with the number of agents. Typically, the heterogeneity distance quantification algorithm outputs a heterogeneity distance matrix for the entire multi-agent system, with a computational complexity of $O(N^2)$. When the number of agents increases significantly, matrix computation becomes costly. However, if only studying heterogeneity between specific agents in the MAS is required, the method remains effective. One only needs to remove data from other agents during CVAE training and sampling computation.

Additionally, the practical algorithms for heterogeneity quantification are built on the assumption that agent-related variables are vectors. If certain agent variables, such as observation inputs, are multimodal, operations like padding in the proposed algorithm become difficult to implement. But this does not affect the correctness of the theory. As the relevant theory still holds in this situation, additional tricks are needed for practical calculation implementation.

\section{Broader Impacts}
\label{app:BroaderImpacts}

Our work systematically analyzes heterogeneity in MARL, which has strong correlations with a series of works in MARL. Under our theoretical framework, research on agent policy diversity in MARL can be categorized within the domain of policy heterogeneity. Our work can give a new perspective for studying policy diversity. Our proposed quantification methods can not only help these works with policy evolution analysis but also explain the relationship between policy diversity and agent heterogeneity. Furthermore, our proposed HetDPS, as an application case, can also be classified among parameter sharing-based works.

Additionally, some traditional heterogeneous MARL works can be categorized within environment-related heterogeneity domains. Our quantification and definition methods are orthogonal to these works, which can fully utilize our proposed methodology for further advancement. For instance, observation heterogeneity quantification can be used to enhance agents' ability to aggregate heterogeneous observation information; transition heterogeneity quantification can help design intrinsic rewards to assist heterogeneous multi-agents in learning cooperative policies.

In conclusion, our work not only expands the scope of heterogeneity in MARL but also closely connects with many current hot topics, contributing to the further development of these works.

\section{An introduction to POMG}
\label{app:POMG}

Partially Observable Markov Game (POMG) is essentially an extension of Partially Observable Markov Decision Process (POMDP), which in turn extends Markov Decision Process (MDP). MDP~\citep{MDP1,MDP2} is a mathematical framework that describes sequential decision-making by a single agent in a fully observable environment. In an MDP, the agent can fully observe the environment's state, select actions based on the current state, and aim to maximize cumulative rewards. Compared to MDP, the key extension of POMDP~\citep{POMDP1,POMDP2} is the consideration of partial observability, making it suitable for modeling both single-agent partially observable problems~\citep{spaan2012partially} and multi-agent problems~\citep{bernstein2002complexity,oliehoek2016concise}. In multi-agent POMDPs, agents typically operate in a fully cooperative mode, where their rewards are usually team-shared.

\begin{figure*}[ht]
\centering
\includegraphics[width=0.7\textwidth]{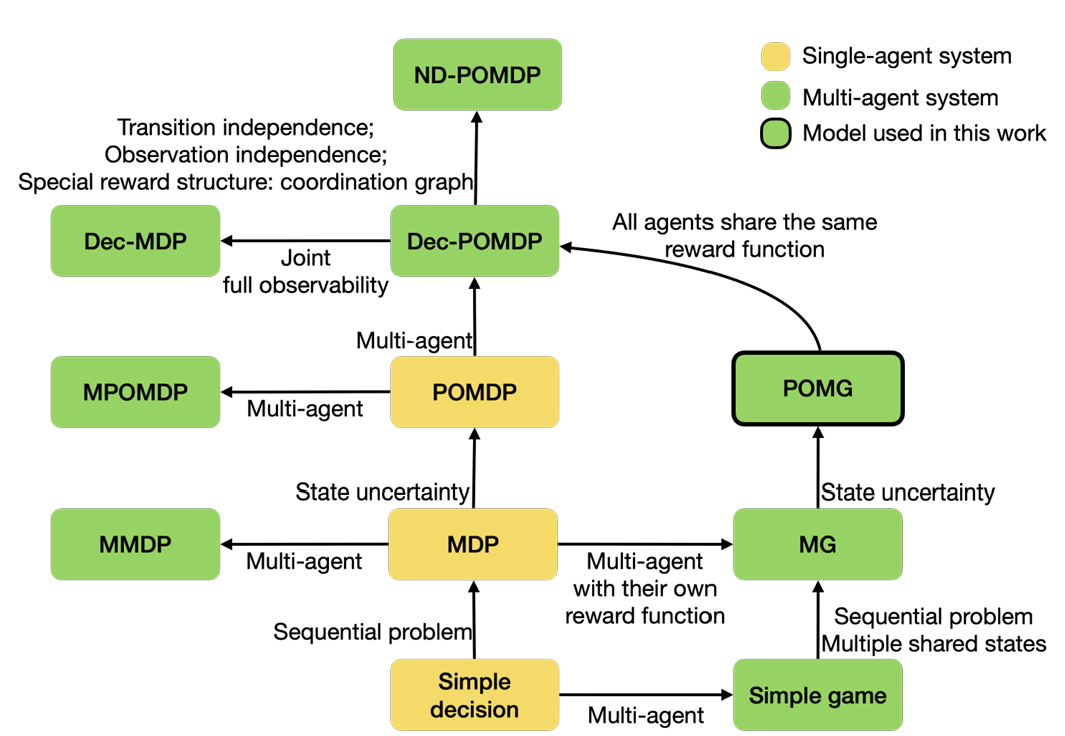}
\caption{Common multi-agent problem formulations~\citep{POMG2}.}
\label{fig:POMGs1}
\Description{Diagram showing common multi-agent problem formulations including MDP, POMDP, and POMG relationships.}
\end{figure*}

\begin{figure*}[ht]
\centering
\includegraphics[width=0.6\textwidth]{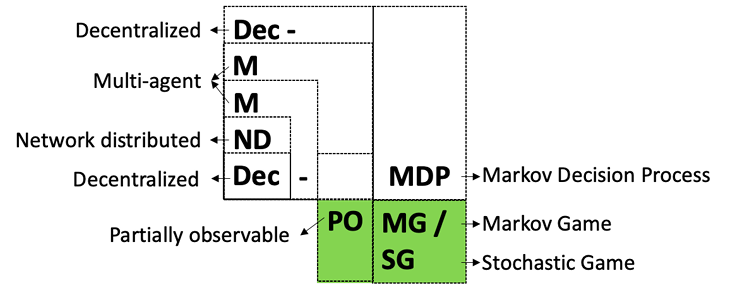}
\caption{Common nomenclature for multi-agent models~\citep{POMGs}.}
\label{fig:POMGs2}
\Description{Diagram illustrating common nomenclature and relationships for multi-agent models in reinforcement learning.}
\end{figure*}

The key extension of POMG over POMDP lies in modeling mixed game relationships among multiple agents. Unlike POMDP, agents in POMG do not share a common reward function; instead, each agent has its own (agent-level) reward function, making POMG more general~\citep{POMGs,MADRLSurvey}. This design enables POMG to handle competitive, cooperative, and mixed interaction scenarios, better reflecting the complexity of real-world multi-agent systems. The logical relationships among Markov decision processes and their variants are illustrated in Figure~\ref{fig:POMGs1} and Figure~\ref{fig:POMGs2}. As shown in these figures, POMG is the most general framework for modeling original problems in the MARL domain. For these reasons, we chose POMG as the foundation for discussing heterogeneity in MARL.

\section{Other potential types of heterogeneity in MARL}
\label{app:More Het}

Benefiting from the reinforcement learning modeling based on POMG, we have clearly defined the boundaries of heterogeneity discussed in this paper. In fact, within the realm of unconventional multi-agent systems, there might be other types of heterogeneity. 

For instance, agents may have different length of decision timesteps, with some agents inclined towards long-term high-level decisions, while others tend to make short-term low-level decisions. Agents may also have different discount factors, some works try to assign varying discount factors to different agents during algorithm training~\citep{role_transfer}, to encourage agents to develop "\textit{myopic}" or "\textit{far-sighted}" policy behaviors, thereby promoting agent cooperation. However, differences in discount factors are more reflective of algorithmic design variations rather than environmental distinctions, and thus fall outside the scope of this paper. Moreover, there may be heterogeneity among agents regarding communication, agents might have different communication channels due to hardware variations. However, the establishment of communication protocols aims to enable agents to receive more information when making decisions, potentially overcoming non-stationarity and partial observability issues~\citep{MADRLSurvey}. These communication messages are essentially mappings of global information processed in the environment, which are then input into the action-related network modules. From this perspective, agent communication can be modeled as a more generalized observation function that maps global information to local observations for agent decision-making, and communication heterogeneity can be categorized under observation heterogeneity. From a learning perspective, agents might also have heterogeneous available knowledge, such as differences in initial basic policies or variations in supplementary knowledge accessible during execution phase. Moreover, heterogeneity might extend beyond abstract issues, including computational resource differences among agents during learning.

Overall, even from the perspective of multi-agent reinforcement learning, heterogeneity in multi-agent systems remains a domain with extensive discussion space, warranting further subsequent research.

\section{Properties of heterogeneity distance}
\label{app:Properties of heterogeneity distance}

\noindent\textcolor{blue!70!black}{\textbf{\large Recap.}} The heterogeneity distance between two agents in Section~\ref{sec:Quantifying} can be computed as follows:

\begin{equation}
    d_{ij}^F = \int_{x \in X} D[F_i(\cdot|x), F_j(\cdot|x)] \cdot p(x) \, dx,
\end{equation}
where $ X $ is the space of independent variables, $ p(x) $ is the probability density function, and $ D [\cdot, \cdot] $ is a measure that quantifies the difference between distributions.

\noindent\textcolor{red!70!black}{\textbf{\large Proposition 1.}} (\textit{Properties of Heterogeneity Distance})
\ding{192} \textcolor{blue!70!black}{\textbf{Symmetry}}: $ d^F_{ij} = d^F_{ji} $;
\ding{193} \textcolor{green!60!black}{\textbf{Non-negativity}}: $ d^F_{ij} \geq 0 $;
\ding{194} \textcolor{orange!70!black}{\textbf{Identity of indiscernibles}}: $ d^F_{ij} = 0 $ if and only if agents $ i $ and $ j $ are $ F $-homogeneous;
\ding{195} \textcolor{purple!70!black}{\textbf{Triangle inequality}}: $ d^F_{ij} \leq d^F_{ik} + d^F_{kj} $ $(i,j,k \in N)$.
This proposition holds as long as the measure $ D $ satisfies Property \ding{192}\ding{193}\ding{194}\ding{195}.

\noindent\textcolor{green!60!black}{\textbf{\large Proof.}} It can be proven that when $D$ satisfies Property \ding{192}\ding{193}\ding{194}\ding{195}, heterogeneity distance  also satisfies Property \ding{192}\ding{193}\ding{194}\ding{195}.

\noindent\textbf{\textcolor{blue!60!black}{1) Proof of Symmetry:}} 
\begin{equation}
\begin{aligned}
    d_{ij}^F &= \int_{x \in X} D\left[F_i(\cdot|x), F_j(\cdot|x)\right] \cdot p(x) d x \\
    &= \int_{x \in X} D\left[F_j(\cdot|x), F_i(\cdot|x)\right] \cdot p(x) d x \\
    &= d_{ji}^F.
\end{aligned}
\end{equation} 

\noindent\textbf{\textcolor{blue!60!black}{2) Proof of Non-negativity:}}  
\begin{equation}
\begin{aligned}
d_{ij}^F &= \int_{x \in X} D\left[F_i(\cdot|x), F_j(\cdot|x)\right] \cdot p(x) d x \\
&\geq \int_{x \in X} 0 \cdot p(x) d x = 0.
\end{aligned}
\end{equation}

\noindent\textbf{\textcolor{blue!60!black}{3) Proof of Identicals of indiscernibility (necessary conditions):}} 

if agent $i$ and agent $j$ are $F$-homogeneous, then we have:
$X^{(i)} = X^{(j)}$, $\forall x \in X=X^{(i)}$, $F_i(\cdot|x) = F_j(\cdot|x)$,
\begin{equation}
\begin{aligned}
d_{ij}^F & = 
\int_{x \in X} D\left[F_i(\cdot|x), F_j(\cdot|x)\right] \cdot p(x) d x \\
&= \int_{x \in X} D\left[F_i(\cdot|x), F_i(\cdot|x)\right] \cdot p(x) d x \\
&=\int_{x \in X} 0 \cdot p(x) d x \\
&= 0.
\end{aligned}
\end{equation}

\noindent\textbf{\textcolor{blue!60!black}{4) Proof of Identicals of indiscernibility (sufficient conditions):}} 
\begin{equation}
\begin{aligned}
d_{ij}^F=0 &\xrightarrow{\text{Prop.\ding{193}}} D\left[F_i(\cdot|x), F_i(\cdot|x)\right] = 0, \\
&\quad \forall x \in X^{(i)} \text{ or } X^{(j)}\\
&\xrightarrow{\text{Prop.\ding{193}of $D$}} F_i(\cdot|x) = F_i(\cdot|x), \\
&\quad \forall x \in X, X = X^{(i)} = X^{(j)}, 
\end{aligned}
\end{equation}
then we have agent $i$ and agent $j$ are $F$-homogeneous.

\noindent\textbf{\textcolor{blue!60!black}{5) Proof of Triangle Inequality:}}  
\begin{equation}
\begin{aligned}
d_{ij}^F &= \int_{x \in X} D\left[F_i(\cdot|x), F_j(\cdot|x)\right] \cdot p(x) \, dx \\
&\leq \int_{x \in X} \left( D\left[F_i(\cdot|x), F_k(\cdot|x)\right] \right. \\
&\quad \left. + D\left[F_k(\cdot|x), F_j(\cdot|x)\right] \right) \cdot p(x) \, dx \\
&= \int_{x \in X} D\left[F_i(\cdot|x), F_k(\cdot|x)\right] \cdot p(x) \, dx \\
&\quad + \int_{x \in X} D\left[F_k(\cdot|x), F_j(\cdot|x)\right] \cdot p(x) \, dx \\
&= d_{ik}^F + d_{kj}^F.
\end{aligned}
\end{equation}

In this paper, we choose the \textcolor{purple!70!black}{\textbf{Wasserstein Distance}}~\citep{WD} as the metric to quantify the distance between distributions, which satisfies the property \ding{192}\ding{193}\ding{194}\ding{195}~\citep{SND}.

\noindent\textcolor{orange!70!black}{\textbf{\large Discussion.}} 
In practical computation, we adopt a representation learning-based approach to find an alternative latent variable distribution $p_i(z|x)$ to replace the original distribution $F_i(y|x)$ for quantification. It can be easily proved that when using latent variable distributions to compute heterogeneous distances, these distances still satisfy properties \ding{192}, \ding{193}, and \ding{195} (following the same proof method as above).

In the model-based case, $p_i(z|x) = f_\phi(y_i,x)$, where $ f_\phi$ represents the encoder of the CVAE. When two agents have the same independent and dependent variables (identical agent functions), their latent variable distributions are also identical. In this case, it is straightforward to prove that property \ding{194} still holds under the model-based case.

In the model-free case, $p_i(z|x) = f_\phi(i,x)$. Due to the lack of an environment model, even agents with identical mappings may learn different representation distributions through their encoders, thus not satisfying property \ding{194}. However, as demonstrated in Section~\ref{sec:case_study}, although we cannot strictly determine agent homogeneity using $d_{ij}^F=0$, the heterogeneity distances measured between homogeneous agents in the model-free case are sufficiently small. Moreover, the model-free manner is adequate to distinguish between homogeneous and heterogeneous agents, and still maintains the ability to quantify the degree of heterogeneity (as shown in Sections~\ref{sec:case_study} and~\ref{sec:exp}).

\section{More details of computing heterogeneity distance}
\label{app:computing heterogeneity distance}

Here, we present five formulas for calculating heterogeneity distances, corresponding to the five types of heterogeneity discussed in this paper.

Regarding \textcolor{blue!70!black}{\textbf{\large Observation Heterogeneity}}, its relevant elements include the agent's observation space and observation function. For two agents $i$ and $j$, let their observation heterogeneity distance be denoted as $d^\Omega_{ij}$. The corresponding calculation formula is:

\begin{equation}
\begin{aligned}
    d_{ij}^\Omega &= \int_{\hat{s} \in \{S^i\}_{i \in N}} D\left[\Omega_i(\cdot|\hat{s}), \right. \\
    &\quad \left. \Omega_j(\cdot|\hat{s})\right] \cdot p(\hat{s}) \, d\hat{s},
\end{aligned}
\label{eq:distance-1}
\end{equation}
where $D [\cdot, \cdot]$ represents a measure of distance between two distributions, and $p(\cdot)$ is the probability density function (this notation applies to subsequent equations). Here, $\hat{s}$ denotes the global state, $\{S^i\}_{i \in N}$ represents the global state space, and $\Omega_i$ and $\Omega_j$ are the observation functions of agents $i$ and $j$, respectively.

Regarding \textcolor{green!60!black}{\textbf{\large Response Transition Heterogeneity}}, its relevant elements include the agent's action space, state space, and global state transition function. For two agents $i$ and $j$, let their response transition heterogeneity distance be denoted as $d^{\mathcal{T}}_{ij}$. The corresponding calculation formula is:

\begin{equation}
\begin{aligned}
    d_{ij}^\mathcal{T} &= \int_{\hat{s} \in \{S^i\}_{i \in N}} \int_{\hat{a} \in \{A^i\}_{i \in N}} \\
    &\quad D\left[\mathcal{T}^i(\cdot|\hat{s},\hat{a}), \mathcal{T}^j(\cdot|\hat{s},\hat{a})\right] \\
    &\quad \cdot p(\hat{s},\hat{a}) \,d\hat{a} d\hat{s},
\end{aligned}
\label{eq:distance-2}
\end{equation}
where $p(\cdot,\cdot)$ represents the joint probability density function. $\hat{s}$ and $\hat{a}$ denote the global state and global action respectively, $\{S^i\}_{i \in N}$ and $\{A^i\}_{i \in N}$ represent the global state space and global action space, and $\mathcal{T}_i$ and $\mathcal{T}_j$ are the local state transition functions of agents $i$ and $j$, respectively.

Regarding \textcolor{orange!70!black}{\textbf{\large Effect Transition Heterogeneity}}, its relevant elements include the agent's action space, state space, and global state transition function. For convenience, we denote $S^{-i} = \times_{k \in N, k \neq i} S^{k} \times S^{E}$ as the joint state space of all agents except agent $i$, $A^{-i} = \times_{k \in N, k \neq i} A^{k}$ as the joint action space of all agents except agent $i$, and $\mathcal{T}^{-i}$ as the collection of state transitions excluding agent $i$. For two agents $i$ and $j$, let their effect transition heterogeneity distance be denoted as $d^{\mathcal{T}^{-}}_{ij}$. The corresponding calculation formula is:

\begin{equation}
\begin{aligned}
    d_{ij}^{\mathcal{T}^{-}} &= 
    \int_{s' \in S^{(-i)}} \int_{s  \in A^{i}} \int_{a' \in A^{(-i)}}
    \int_{a  \in A^{i}} \\
    &\quad D\left[\mathcal{T}^{-i}(\cdot|x), \mathcal{T}^{-j}(\cdot|x)\right]
    \cdot p(x) \, da \, da' \, ds \, ds',
\end{aligned}
\label{eq:distance-3}
\end{equation}

where for convenience, we denote $x=(s',s,a',a)$, and $p$ is the joint probability density function.

The calculation of effect transition heterogeneity distance differs from the previous two types of heterogeneity distances in two significant ways. The first difference lies in its introduction of agent-level elements as variables rather than global variables. When two agents have different agent-level variable spaces, it becomes challenging to calculate the heterogeneity distance under this definition.
The second difference is that it involves a quadruple integral, making its computational complexity much higher than the single or double integrals of the previous two distances.

These two differences make the calculation of effect transition heterogeneity distance more challenging. Fortunately, through our proposed meta-transition model, we can simplify the calculation of effect transition heterogeneity distance to a double integral that only involves the agent's local states and actions. Additionally, the distance measurement through representation learning also reduces the constraints on the similarity of agents' variable spaces. Even when two agents have different variable spaces (for example, one agent's local state space is 10-dimensional while another's is 20-dimensional), we can still process the variable inputs through techniques like padding and then map them to the same dimension using encoder networks. This demonstrates that the approach based on representation learning and meta-transition significantly extends the applicability of heterogeneity distance measurement, which also holds true in the quantification of heterogeneous types discussed below.

Regarding \textcolor{purple!70!black}{\textbf{\large Objective Heterogeneity}}, its relevant element is the agent's reward function. For two agents $i$ and $j$, let their objective heterogeneity distance be denoted as $d^{r}_{ij}$. The corresponding calculation formula is:

\begin{equation}
\begin{aligned}
    d_{ij}^r &= \int_{\hat{s} \in \{S^i\}_{i \in N}} \int_{\hat{a} \in \{A^i\}_{i \in N}} \\
    &\quad D\left[r^i(\cdot|\hat{s},\hat{a}), r^j(\cdot|\hat{s},\hat{a})\right] \\
    &\quad \cdot p(\hat{s},\hat{a}) \,d\hat{a} d\hat{s},
\end{aligned}
\label{eq:distance-4}
\end{equation}
where $p(\cdot,\cdot)$ represents the joint probability density function. $\hat{s}$ and $\hat{a}$ denote the global state and global action respectively, $\{S^i\}_{i \in N}$ and $\{A^i\}_{i \in N}$ represent the global state space and global action space, and $r_i$ and $r_j$ are the reward functions of agents $i$ and $j$, respectively.

Regarding \textcolor{red!70!black}{\textbf{\large Policy Heterogeneity Distance}}, its relevant elements include the agent's observation space, action space, and policy function. For two agents $i$ and $j$, let their policy heterogeneous distance be denoted as $d^\pi_{ij}$. The corresponding calculation formula is:

\begin{equation}
d_{ij}^\pi = \int_{o \in O^i} D\left[\pi_i(\cdot|o), \pi_j(\cdot|o)\right] \cdot p(o) \, do,
\label{eq:distance-5}
\end{equation}

where $D [\cdot, \cdot]$ represents a measure of distance between two distributions, and $p(\cdot)$ is the probability density function. Here, $o$ denotes the observation, $O^i$ represents the observation space, and $\pi_i$ and $\pi_j$ are the policy functions of agents $i$ and $j$, respectively.

\section{Meta-Transition and its Heterogeneity Distance} \label{app:MetaTransition}

To quantify an agent's comprehensive heterogeneity, we introduce the concept of \textcolor{blue!70!black}{\textbf{meta-transition}}. \textcolor{blue!70!black}{\textbf{Meta-transition}} is a modeling approach that explores an agent's own attributes from its perspective. Our goal is to quantify an agent's comprehensive heterogeneity using only the agent's local information (as global information is typically difficult to obtain in practical MARL scenarios).

Based on this, we provide the definition of meta-transition. Let the meta-transition of agent $i$ be denoted as \textcolor{purple!70!black}{\textbf{$M_i$}}. It is a mapping \textcolor{purple!70!black}{\textbf{$M_i: S_i \times A_i \rightarrow S_i \times R \times \Omega_i$}}. At time step $t$, the \textcolor{green!60!black}{\textbf{inputs}} of meta-transition are the agent's local state $s_t^i$ and local action $a_t^i$, and the \textcolor{orange!70!black}{\textbf{outputs}} are the next time step's local state $s_{t+1}^i$, the next time step's local observation $o_{t+1}^i$, and the current time step's reward $r_t^i$ based on the state and action.

We explain why the above relationship can reflect all agent-level elements in POMG. The input local state and local action of meta-transition actually correspond to the \textcolor{green!60!black}{\textbf{inverse mapping}} to the global state and global action. This inverse mapping potentially restores the local state and action to global information, and then obtains the next time step's global state according to the global state transition function, which is mapped to local observation through the observation function. Therefore, this process reflects the agent's \textcolor{orange!70!black}{\textbf{effect transition heterogeneity}} and \textcolor{blue!70!black}{\textbf{observation heterogeneity}}. Additionally, the potential global state and global action also determine the agent's local state and corresponding reward at the next time step, which reflect the agent's \textcolor{green!60!black}{\textbf{response transition heterogeneity}} and \textcolor{red!70!black}{\textbf{objective heterogeneity}}, respectively.

It is worth noting that meta-transition is not a function that actually exists in POMG, but an \textcolor{red!70!black}{\textbf{implicitly defined mapping}}. We aim to quantify this mapping difference to capture the agent's comprehensive heterogeneity. Therefore, meta-transition heterogeneity is quantified in a \textcolor{blue!70!black}{\textbf{model-free manner}}.

Moreover, meta-transition is not limited to the aforementioned form. It can be transformed into different forms according to the modular settings of independent and dependent variables. For example, by removing the agent's reward, meta-transition can reflect the agent's \textcolor{blue!70!black}{\textbf{observation heterogeneity}}, \textcolor{green!60!black}{\textbf{response transition heterogeneity}}, and \textcolor{orange!70!black}{\textbf{effect transition heterogeneity}}.

After determining the input and output of meta-transition, the relevant heterogeneity distance can be calculated using the same model-free method as before. Since meta-transition involves multiple variables, and the dimensions between these variables may differ significantly (for example, the dimension of reward is 1, while the dimension of observation might be 100), directly fitting with deep networks may struggle to capture information corresponding to low-dimensional variables. We address this issue through a \textcolor{purple!70!black}{\textbf{dimension replication trick}}. In practice, we typically replicate the reward dimension to be similar to the dimensions of observation or action, ensuring that the autoencoder network can capture information related to objective heterogeneity during learning.

\section{Derivation of ELBO} 
\label{app:CVAE}
\textcolor{blue!70!black}{\textbf{The Evidence Lower Bound (ELBO)}} of the likelihood can be derived as follows:

\begin{equation}
\begin{aligned}
\log p(y|x) 
& = \log \int p(y,z|x) d z   
\quad \quad \quad \quad \quad \quad \ \textcolor{orange!70!black}{\textbf{ (a) }} \\
& = \log \int \frac{ p(y,z|x) f_{\phi}(z|y,x)}{f_{\phi}(z|y,x)}  d z \\
&\quad \quad  \textcolor{orange!70!black}{\textbf{ (b) }} \\
& = \log \mathbb{E}_{f_{\phi}(z|y,x)}
\left[
\frac{p(y,z|x)}{f_{\phi}(z|y,x)}
\right] \\
&\quad \quad \textcolor{orange!70!black}{\textbf{ (c) }} \\
& \geq \mathbb{E}_{f_{\phi}(z|y,x)} 
\left[ \log
\frac{ p(y,z|x)}{f_{\phi}(z|y,x)}
\right] \\
&\quad \quad \textcolor{orange!70!black}{\textbf{ (d) }} \\
& = ELBO_{\text{model-based}},
\end{aligned}
\end{equation}
where $f_{\phi}(z|y,x)$ represents the \textcolor{green!60!black}{\textbf{posterior probability distribution}} of the latent variable generated by the encoder, and $ p(y,z|x)$ denotes a \textcolor{green!60!black}{\textbf{joint probability distribution}} concerning the customized feature and latent variable, conditioned on $o$. Throughout the derivation of the formula, \textcolor{orange!70!black}{\textbf{(a)}} employs the properties of the joint probability distribution, \textcolor{orange!70!black}{\textbf{(b)}} multiplies both numerator and denominator by $f_{\phi}(z|y,x)$, \textcolor{orange!70!black}{\textbf{(c)}} applies the definition of mathematical expectation, and \textcolor{orange!70!black}{\textbf{(d)}} invokes the Jensen's inequality.

Considering that the ELBO includes an unknown joint probability distribution, we can further decompose it by using the posterior probability distributions from the encoder and decoder:

\begin{equation}
\begin{aligned}
ELBO_{\text{model-based}}
& = \mathbb{E}_{f_{\phi}(z|y,x)} 
\left[ \log
\frac{ p(y,z|x)}{f_{\phi}(z|y,x)}
\right] \\
& = 
\mathbb{E}_{f_{\phi}(z|y,x)} 
\left[ \log
\frac{g_\omega(c|z,x) p(z|x)}{f_{\phi}(z|y,x)}
\right] \\
&\quad \quad \textcolor{orange!70!black}{\textbf{ (a)}} \\
& =
\mathbb{E}_{f_{\phi}(z|y,x)} 
\left[ \log
g_\omega(c|z,x)
\right] \\
& \quad +
\mathbb{E}_{f_{\phi}(z|y,x)} 
\left[ \log
\frac{ p(z|x)}{f_{\phi}(z|y,x)}
\right] \\
&\quad \textcolor{orange!70!black}{\textbf{ (b) }} \\
& =\underbrace{\mathbb{E}_{f_{\phi}(z|y,x)} \left[\log
g_\omega(c|z,x)\right]}_{\textcolor{blue!70!black}{\textbf{reconstruction term}} } \\
&\quad -
\underbrace{D_{\mathrm{KL}}\left[f_{\phi}(z|y,x)
 \| p(z|x)\right]}_{\textcolor{red!70!black}{\textbf{prior matching term}} }, \\
&\quad \textcolor{orange!70!black}{\textbf{ (c) }} \\
\end{aligned}
\end{equation}
where $f_{\phi}(z|y,x)$ and $g_\omega(c|z,x)$ are the \textcolor{green!60!black}{\textbf{posteriors}} from the encoder and decoder, respectively. The conditional joint probability distribution $ p(y,z|x)$ is a imaginary construct in mathematical terms and lacks practical significance. It can be formulated using the probability chain rule, constructed from the posterior distribution of the customized feature and the prior distribution of the latent variable (step \textcolor{orange!70!black}{\textbf{(a)}}). Step \textcolor{orange!70!black}{\textbf{(b)}} decomposes the expectation, and step \textcolor{orange!70!black}{\textbf{(c)}} applies the definition of the KL divergence.

Thus, the ELBO can be decomposed into a \textcolor{blue!70!black}{\textbf{reconstruction term}} of the customized feature, and a \textcolor{red!70!black}{\textbf{prior matching term}} of the posterior and the prior. By maximizing the ELBO, the reconstruction likelihood can be maximized while minimizing the KL divergence between the posterior and the prior. In the model-free case, the same approach can be used to derive the ELBO and corresponding loss function.

\section{Details of HetDPS}
\label{app:HetDPS}

\textcolor{blue!70!black}{\textbf{\large HetDPS}} is a novel algorithm designed to efficiently manage the allocation of neural network parameters across multiple agents in MARL. This algorithm leverages the \textcolor{purple!70!black}{\textbf{Wasserstein distance matrix}} to cluster agents based on their similarities, and subsequently assigns them to suitable neural networks. The pseudocode of HetDPS is shown in Algorithm~\ref{alg:dynamic_parameter_sharing}.

The algorithm begins by computing the \textcolor{green!60!black}{\textbf{affinity matrix}} from the Wasserstein distance matrix, which is then used as input to the \textcolor{green!60!black}{\textbf{Affinity Propagation clustering algorithm}}. This process yields a new set of cluster assignments for the agents. If it is the first time the algorithm is executed, the cluster assignments are directly used as network assignments.

In subsequent iterations, the algorithm compares the new cluster assignments with the previous ones to determine the optimal network assignments. This is achieved by constructing an \textcolor{orange!70!black}{\textbf{overlap matrix}} that captures the similarity between the old and new cluster assignments. Based on the number of old and new clusters, the algorithm handles three distinct cases:

\textcolor{blue!70!black}{\textbf{1. Equal number of old and new clusters:}} In this scenario, the algorithm establishes a one-to-one mapping between the old and new clusters using the \textcolor{red!70!black}{\textbf{Hungarian algorithm}}. It then constructs a mapping from old clusters to networks and assigns each agent to a network based on its new cluster assignment.

\textcolor{green!60!black}{\textbf{2. More new clusters than old clusters:}} When the number of new clusters exceeds the number of old clusters, the algorithm handles \textcolor{red!70!black}{\textbf{network splitting}}. It uses the Hungarian algorithm to find the best matching between old and new clusters and establishes a mapping from new clusters to old clusters. For new clusters without a clear match, the algorithm either finds the most similar old cluster or identifies the closest network. It then executes a splitting operation to copy parameters from the source network to the new network.

\textcolor{orange!70!black}{\textbf{3. More old clusters than new clusters:}} In this case, the algorithm handles \textcolor{red!70!black}{\textbf{network merging}}. It uses the Hungarian algorithm to find the best matching between old and new clusters and establishes a mapping from old clusters to new clusters. For each new cluster, it identifies the networks to be merged and executes a merging operation based on the specified merge mode (\textcolor{purple!70!black}{majority, random, average, or weighted}). The algorithm then assigns each agent to a network based on its new cluster assignment.

HetDPS offers a flexible and efficient approach to managing neural network parameters in multi-agent systems. By dynamically adjusting network assignments based on agent similarities, the algorithm enables effective parameter sharing and reduces the need for redundant computations.

\begin{algorithm}[ht]
\caption{\textbf{HetDPS}}
\label{alg:dynamic_parameter_sharing}
\begin{algorithmic}[1]
\State \textcolor{blue!70!black}{\textbf{Initialize}} policies and parameter sharing paradigm
\For{\textcolor{green!60!black}{\texttt{episode}} = 1 to \textcolor{green!60!black}{\texttt{maxEpisodes}}}
    \State \textcolor{green!60!black}{Interact} with environment to collect data
    \State \textcolor{green!60!black}{Add} data to reinforcement learning (RL) sample pool
    \State \textcolor{green!60!black}{Add} data to heterogeneity distance sample pool
    \If{\textcolor{orange!70!black}{\texttt{episode}} \% \textcolor{orange!70!black}{\texttt{trainingPeriod}} = 0}
        \State \textcolor{orange!70!black}{\textbf{Update}} policies using RL sample pool
    \EndIf
    \If{\textcolor{purple!70!black}{\texttt{episode}} \% \textcolor{purple!70!black}{\texttt{quantizationPeriod}} = 0}
    \State \textcolor{purple!70!black}{\textbf{Compute}} heterogeneity distance matrix $D$ (Section~\ref{sec:Quantifying})
        \State \textcolor{purple!70!black}{\textbf{Cluster}} agents using Affinity Propagation on $D$
        \If{no previous clustering exists}
            \State \textcolor{blue!60!black}{Assign} networks to agents based on clusters
            \State \textcolor{blue!60!black}{Copy} network parameters as needed
        \Else
            \State \textcolor{blue!60!black}{Compute} maximum overlap matching between current and previous clusters
            \If{number of clusters unchanged}
                \State \textcolor{teal!70!black}{Map} new clusters to previous networks
            \ElsIf{new clusters $>$ previous clusters}
                \State \textcolor{red!70!black}{\textbf{Split}} networks: copy parameters for unmatched clusters
            \Else
                \State \textcolor{red!70!black}{\textbf{Merge}} networks: combine parameters based on merge mode
            \EndIf
            \State \textcolor{blue!60!black}{Assign} networks to agents
        \EndIf
    \EndIf
\EndFor
\end{algorithmic}
\end{algorithm}